\RequirePackage{fix-cm}
 \documentclass[smallextended]{svjour3}       
\smartqed  
\usepackage{graphicx}
\usepackage{latexsym}
\begin{document}

\title{Transverse Single Spin Asymmetry in $J/\psi$ Production
}


\author{Bipin Sonawane        \and
        Anuradha Misra \and
        Siddhesh Padval \and
        Vaibhav Rawoot
}


\institute{Bipin Sonawane \at
              Department of Physics, University of Mumbai, Vidyanagari, Santacruz-East. Mumbai 400 098. \\
              Tel.: +91-022-652 6250\\
              Fax: +91-022-652 2222\\
              \email{bipin.sonawane@physics.mu.ac.in}           
           \and
           Anuradha Misra, Siddhesh Padval and Vaibhav Rawoot \at
               Department of Physics, University of Mumbai, Santacruz-East. Mumbai 400 098.
}

\date{Received: date / Accepted: date}

\maketitle

\begin{abstract}
We estimate transverse single spin asymmetry (TSSA) in electroproduction of $J/\psi$ for J-Lab and EIC energies. We present estimates of TSSAs in $J/\psi$ production within generalized parton model (GPM) using recent parametrizations of gluon Sivers function (GSF) and compare the results obtained using color singlet model (CSM) 
with those obtained using color evaporation model (CEM) of quarkonium production.
\keywords{TSSA \and GSF \and GPM \and CSM \and CEM }
\end{abstract}

\section{Introduction}
\label{intro}
The intrinsic dynamics of partons within proton with finite transverse momentum has been studied within generalized parton model (GPM). GPM is an extension of the collinear factorization theorem taking into account intrinsic transverse momentum of parton.
The collinear parton distribution functions and fragmentation functions with inclusion of intrinsic transverse momentum are called transverse momentum dependent PDFs (TMD-PDFs) and transverse momentum dependent fragmentation functions (TMD-FFs), collectively referred to as TMDs.
The gluon Sivers Function (GSF) \cite{Sivers:1989cc,Sivers:1990fh} is an important TMD which gives the the probability of finding unpolarized quarks inside the transversely polarized proton. The correlation between proton spin and parton transverse momentum in Sivers function enables us to understand the spin dynamics of proton.

TSSA for inclusive process $ A^\uparrow + B \rightarrow C + X $ is defined as $A_N = \frac{d\sigma^\uparrow  -  d\sigma^\downarrow}{d\sigma^\uparrow  + d\sigma^\downarrow}$,
where $ d\sigma^\uparrow $ and $ d\sigma^\downarrow $ represent the differential cross section for scattering of a transversely polarized hadron A off an unpolarized hadron (or lepton) B with A being upwards (downwards) polarized with respect to the production plane.

In this work, we have estimated  TSSA  in low virtuality electroproduction of $J/\psi$ production using recent parametrizations of the GSF. This process is a clean probe of the GSF as the only partonic process involved at leading order is $\gamma g \to c\bar c$. The comparison of asymmetry obtained using the color evaporation model (CEM) and the color singlet model (CSM) of $J/\psi$ production gives a good opportunity to test the production mechanism of $J/\psi$ .

\section{Formalism}
\subsection{Color Evaporation Model}
 Considering a generalization of the CEM~\cite{GayDucati:1999kh} expression, which takes into account 
 the transverse momentum dependence of Weizsacker-William (WW) function 
 and parton distribution function, the expressions for numerator and 
 denominator of asymmetry are~\cite{Godbole:2012bx}
\begin{eqnarray}
{d \sigma^{\uparrow}} - {d \sigma^{\downarrow}}= F_{J/\psi} \int_{4m_{c}^2}^{4m_{D}^2}dM^2 \int d^2 {\bf k}_{\perp g} \Delta^N f_{g/p^{\uparrow}} (x_g, {\bf k}_{\perp g}) 
f_{\gamma / e }(x_\gamma, {\bf k}_{\perp \gamma}) \hat\sigma_0
\end{eqnarray} 

\begin{eqnarray} 
{d \sigma^{\uparrow}} + {d \sigma^{\downarrow}}= 2 F_{J/\psi} \int_{4m_{c}^2}^{4m_{D}^2}dM^2 \int d^2 {\bf k}_{\perp g}  f_{g/p^{\uparrow}} (x_g, {\bf k}_{\perp g})  f_{\gamma / e }(x_\gamma, {\bf k}_{\perp \gamma}) \hat\sigma_0
\end{eqnarray} 
where, $d\sigma^{\uparrow (\downarrow)} = \frac{d \sigma^{\uparrow(\downarrow)}}{dy d^2{\bf q}_T}$ and ${\bf k}_{\perp \gamma}={\bf q}_T - {\bf k}_{\perp g}$. Parameter $F_{J/\psi}$ gives the probability of $J/\psi$ production below $D \bar D$ threshold and ${\hat\sigma_0(M^2)}$ is partonic cross-section for sub-process $ \gamma g \to c\bar c$. Unpolarized TMD-PDF is assumed to have a Gaussian form 
\begin{equation}
f_{g/p}(x,k_\perp;Q)=f_{g/p}(x,Q)\frac{1}{\pi\langle k_\perp^2\rangle}e^{-k_\perp^2/\langle k_\perp^2\rangle}.\end{equation}
and  the parametrization used for the GSF is~\cite{DAlesio:2015fwo}
\begin{equation}
\Delta^N f_{g/p^\uparrow}(x,k_\perp;Q)=2\mathcal{N}_{g}(x)f_{g/p}(x,Q)h(k_\perp)
\frac{e^{-k^2_\perp/\langle k_\perp^2\rangle}}{\pi \langle k_\perp^2\rangle},   \nonumber
\end{equation}
with
\begin{equation}
\mathcal{N}_g(x)=N_g
x^{\alpha_g}(1-x)^{\beta_g}\frac{(\alpha_g+\beta_g)^{\alpha_g+\beta_g}}{\alpha_g^{\alpha_g} \beta_g^{\beta_g}} 
\,{,}\, 
h(k_\perp)=\sqrt{2e}\frac{k_\perp}{M_1}e^{-k_\perp^2/M_1^2},
\end{equation}
where $N_g$, $\alpha_g$, $\beta_g$ and $M_1$ are the best fit parameters. 

\begin{table}[ht]
\centering
\caption{DMP-SIDIS fit parameters from Ref.~\cite{DAlesio:2015fwo}. Here $\rho=M_1^2/(\langle k_\perp^2\rangle+M_1^2)$. Best fit parameters of quark Sivers function used in BV(A) and BV(B) parametrization from Ref.~\cite{Anselmino:2011gs}.}
\begin{tabular}{|l|l|l|l|l|l|l|}
\hline
DMP-SIDIS1 & \multicolumn{2}{l|}{$N_g=0.65$} & $\alpha_g=2.8$ & $\beta_g=2.8$ & $\rho=0.687$ & {$\langle k^2_\perp\rangle=0.25$ GeV$^2$} \\ \cline{1-6}
DMP-SIDIS2 & \multicolumn{2}{l|}{$N_g=0.05$} & $\alpha_g=0.8$ & $\beta_g=1.4$ & $\rho=0.576$ &                                                        \\ \cline{1-7}
\end{tabular}
\begin{tabular}{|l|l|l|l|l|}
\hline
BV (A) and BV (B)&$N_u=0.4$   & $\alpha_u=0.35$ & $\beta_u=2.6$ & $\langle k_\perp^2\rangle=0.25$ GeV$^2$ \\ 
\cline{2-5}
 &$N_d=-0.97$ & $\alpha_d=0.44$ & $\beta_d=0.9$  & $M_1^2=0.19$                                  \\ \hline
\end{tabular}
\label{SIDIS-gluon-fits}
\end{table}
Best fit parameter sets used are given in Table~\ref{SIDIS-gluon-fits} and shall be referred to as  Boer$-$Vogelsang, (BV(A) and BV(B))~\cite{Anselmino:2011gs} and  D'Alesio-Murgia-Pisano
(DMP-SIDIS1 and DMP-SIDIS2)~\cite{DAlesio:2015fwo} parameter sets.  
The comparison of TSSA obtained using BV(A) and  BV(B) parametrization 
with those obtained using  DMP-SIDIS1 and DMP-SIDIS2 parameter 
sets of the GSF is given in Fig.~\ref{p0}. Note that BV(A) and BV(B) parametrization for GSF corresponds to 
$\mathcal{N}_g = \frac{\mathcal{N}_u + \mathcal{N}_d}{2}$ and $\mathcal{N}_g = \mathcal{N}_d$ respectively~\cite{Boer:2003tx}.

\subsection{Color Singlet Model}
\begin{figure}[ht]
\begin{center}
\includegraphics[width=0.38\linewidth,angle=0]{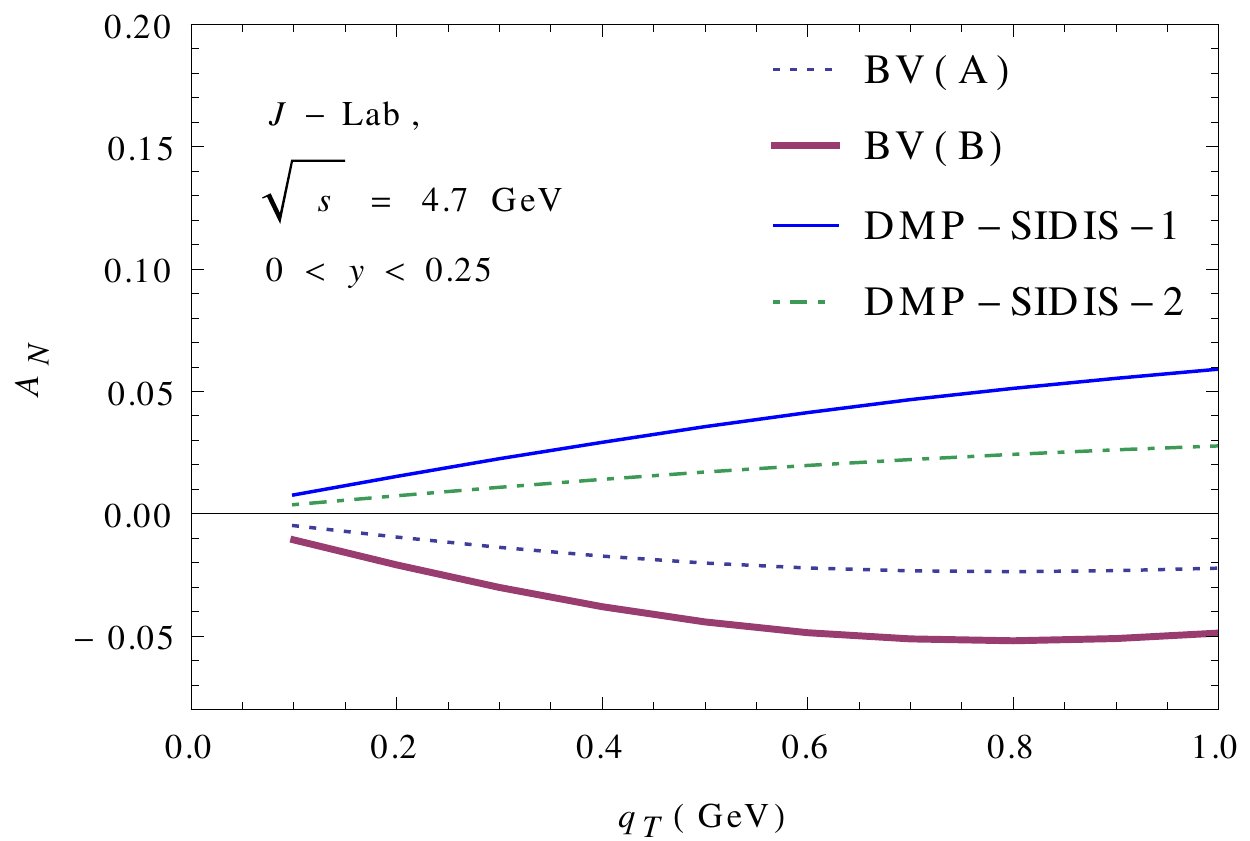}
\includegraphics[width=0.38\linewidth,angle=0]{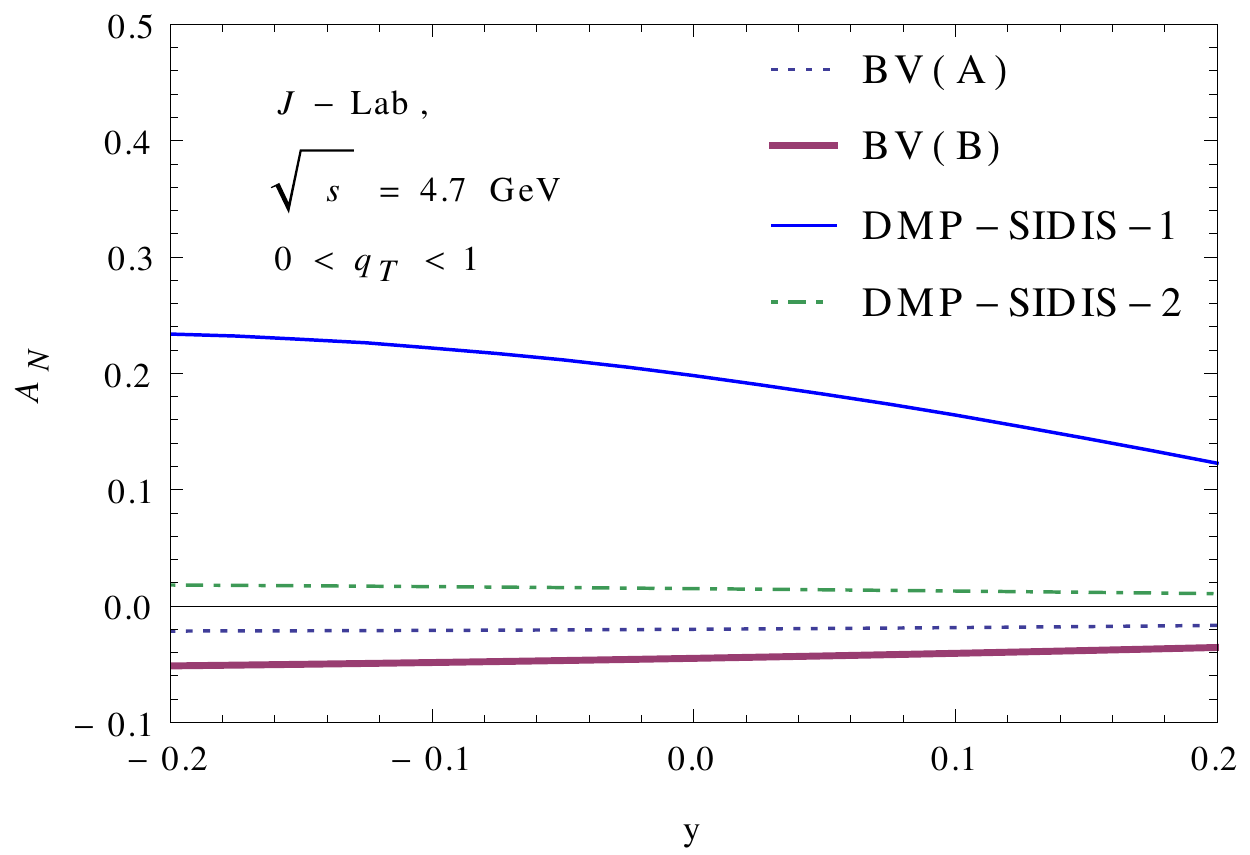}
\caption {Predictions of $q_T $-asymmetry at J-Lab ($\sqrt{s}=$ 4.7 GeV) energy using DGLAP evolved densities with BV(A), BV(B), DMP-SIDIS1 and DMP-SIDIS2 parametrization of GSF. 
}
\label{p0}
\end{center}
\end{figure}

We generalize the photoproduction cross section of $J/\psi$ in CSM \cite{Berger:1980ni,Baier:1981zz,Baier:1982yr}, which is based on the assumption that $J/\psi$ is produced via photon gluon scattering in color singlet state described by a nonrelativistic wave function and employ TMD generalization of the WW approximation to 
 express the electroproduction cross section of $J/\psi$ as 
\begin{eqnarray}
E_{J/\psi}\frac{d\sigma^{e p \rightarrow J/\psi X}}{d^3\textbf{q}_{J/\psi}}=
\int dx_{g}dx_{\gamma}d^2\textbf{k}_{\perp g}d^2\textbf{k}_{\perp \gamma}
\frac{\hat{s}}{x_{g}x_{\gamma}s}\frac{\hat{s}}{\pi} \nonumber \\
f_{\gamma/e}(x_g,\textbf{k}_{\perp g})f_{g/p}(x_\gamma,\textbf{k}_{\perp \gamma})  
\frac{d\hat\sigma^{\gamma g \rightarrow J/\psi g}}{d\hat t}
\delta(\hat{s}+\hat{t}+\hat{u}-m_{J}^{2}).
\end{eqnarray}
where the partonic cross section is
\begin{eqnarray}
\frac{d\hat\sigma^{\gamma g \rightarrow J/\psi g}}{d\hat{t}}=\frac{\alpha_{s}^{2}m_{J}^{3}\Gamma_{ee}^{J}}{\alpha}\frac{8 \pi}{3\hat{s}^{2}}\bigg[\frac{\hat{s}^{2}(\hat{s}-m_{J}^{2})^{2}+\hat{t}^{2}(\hat{t}-m_{J}^{2})^{2}+\hat{u}^{2}(\hat{u}-m_{J}^{2})^{2}}{(\hat{s}-m_{J}^{2})^{2}(\hat{t}-m_{J}^{2})^{2}(\hat{u}-m_{J}^{2})^{2}}\bigg].
\end{eqnarray}
$\Gamma_{ee}^{J}$ is electronic width of $J/\psi$ which is related to $|R_{s}(0)|^2$, $R_{s}(0)$ being the radial wave function at the origin~\cite{Baier:1981zz}. 
We give the results of asymmetry at fix $x_{F}$=0.419 and $\sqrt{s}$=140 GeV in Fig.~\ref{p0}.
A similar study of hadroproduction of $J/\psi$ in CSM has been given recently in Ref.~\cite{DAlesio:2017rzj}.
\begin{figure}[ht]
\begin{center}
\includegraphics[width=0.38\linewidth,angle=0]{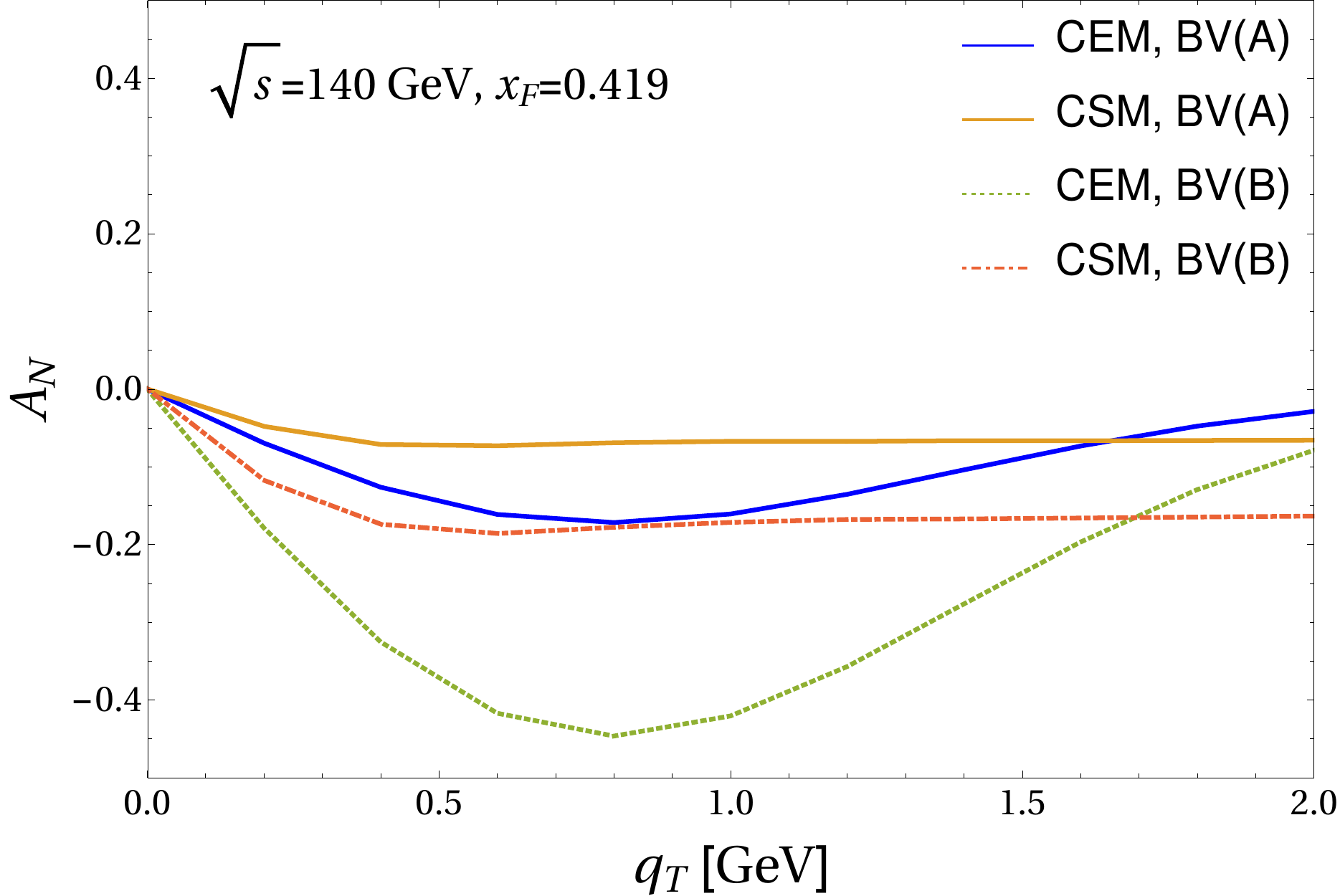}
\includegraphics[width=0.38\linewidth,angle=0]{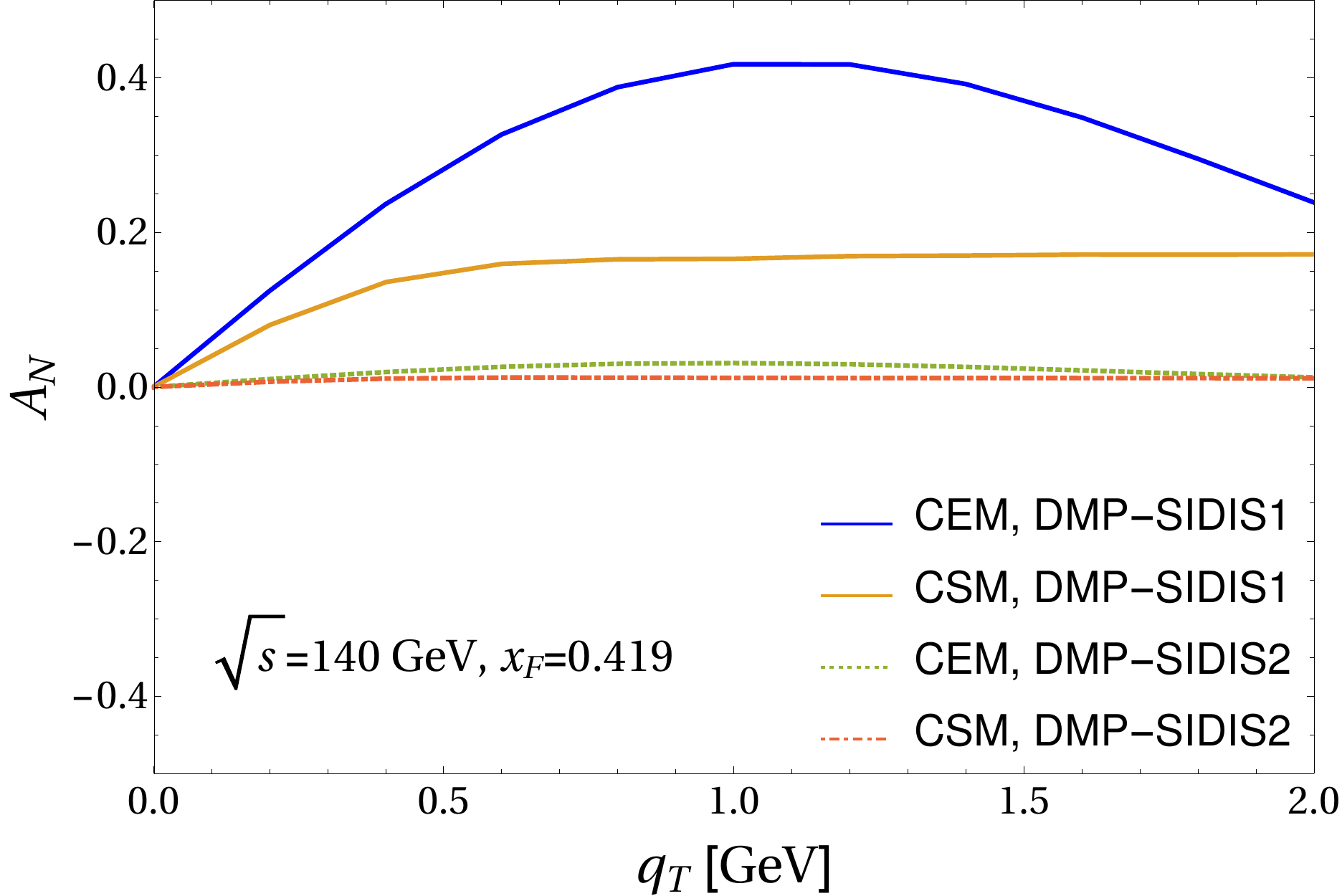}
\caption {Comparison of $q_T $-asymmetry in $e + p^\uparrow \rightarrow e+J/\psi +X$ at $\sqrt{s}=$ 140 GeV and $x_{F}$=0.419 using color evaporation model and color singlet model. Scale is chosen as transverse mass of $J/\psi$ in CSM.
}
\label{p1}
\end{center}
\end{figure}

\section{Conclusions}
In this work, we have presented asymmetry in  $e + p^\uparrow \rightarrow e+J/\psi +X$ process at J-Lab and EIC energy scales using color evaporation model and color singlet model of $J/\psi$ production. We compare our earlier predictions of asymmetry obtained using BV parametrizations with DMP-SIDIS parametrizations. We compare asymmetries obtained using color singlet model and color evaporation model of $J/\psi$ production. Our preliminary results show a sizable asymmetry resulting from recently extracted GSF for both models. A more detailed study taking into account TMD evolution is under progress~\cite{A. Misra : Misra:2018cs}. The comparison of asymmetry obtained using CEM and CSM of $J/\psi$ production could be an important test to understand the production mechanism of heavy quarkonium.
%
 
\end{document}